\documentclass[pre,twocolumn,superscriptaddress,showpacs]{revtex4}

\usepackage{graphicx}
\usepackage{verbatim}
\usepackage{amsmath,amssymb}

\newcommand{\wl}{Wang-Landau}

\begin{document}

\title{Overcoming the critical slowing down of flat-histogram Monte Carlo
simulations: Cluster updates and optimized broad-histogram ensembles}

\author{Yong Wu}
\affiliation{Department of Physics, University of
Massachusetts, Amherst, MA 01003-3720}

\author{Mathias K\"orner}
\affiliation{Theoretische Physik, ETH Z\"urich, CH-8093 Z\"urich,
Switzerland}

\author{Louis Colonna-Romano}
\altaffiliation[Present address: ]{Department of Physics, WPI, Worcester, 
MA 01609-2280}
\affiliation{Department of Physics, Clark University, Worcester, 
MA 01610-1477}

\author{Simon Trebst}
\affiliation{Theoretische Physik, ETH Z\"urich, CH-8093 Z\"urich,
Switzerland}
\affiliation{Computational Laboratory, Eidgen\"ossische Technische
Hochschule, Z\"urich, CH-8092 Z\"urich, Switzerland}

\author{Harvey Gould}
\affiliation{Department of Physics, Clark University, Worcester, 
MA 01610-1477}

\author{Jonathan Machta}
\affiliation{Department of Physics, University of
Massachusetts, Amherst, MA 01003-3720}

\author{Matthias Troyer}
\affiliation{Theoretische Physik, ETH Z\"urich, CH-8093 Z\"urich, Switzerland}

\date{\today}

\begin{abstract}
We study the performance of Monte Carlo simulations that sample a broad
histogram in energy by determining the mean first-passage time to span
the entire energy space of
$d$-dimensional ferromagnetic Ising/Potts models. We first show that
flat-histogram Monte Carlo methods with single-spin flip updates such as the
Wang-Landau algorithm or the multicanonical method perform sub-optimally in
comparison to an unbiased Markovian random walk in energy space. For the
$d=1,2,3$ Ising model, the mean first-passage time $\tau$
scales with the number of spins $N=L^d$ as $\tau
\propto N^2L^z$. The critical exponent $z$ is found to decrease as the
dimensionality 
$d$ is increased.
In the mean-field limit of infinite dimensions we find that
$z$ vanishes up to logarithmic corrections. We then demonstrate how the
slowdown characterized by $z>0$ for finite $d$ can be overcome by two
complementary approaches -- cluster dynamics in connection with Wang-Landau
sampling and the recently developed ensemble optimization technique.
Both approaches are found to improve the random walk in energy space so that
$\tau \propto N^2$ up to logarithmic corrections for the $d=1,2$ Ising model.
\end{abstract}


\maketitle

\section{Introduction}

Recently, Wang and Landau~\cite{wl1,wl2} introduced a Monte Carlo
(MC) algorithm that simulates a biased random walk in energy space,
systematically estimates the density of states, $g(E)$, and iteratively
samples a flat histogram in energy. The bias depends on the
total energy $E$ of a configuration and is defined by a statistical ensemble
with weights $w(E)= 1/g(E)$. The idea is that the probability distribution
of the energy, $p(E) = w(E)g(E)$, will eventually become constant, producing
a flat energy histogram. The algorithm has been successfully applied to a
wide range of systems~\cite{gubernatis}.

One measure of the performance of the \wl\ algorithm and other broad-histogram
algorithms is the mean first-passage time $\tau$, which we define to
be the number of MC steps per spin for the system to go from the configuration 
of lowest energy to the configuration of highest energy. This time has been
called the tunneling time or round-trip time in earlier
work~\cite{tt,EnsembleOptimization}. This time is the relevant scale for
sampling the entire energy space. Because the number of possible energy
values in the Ising model scales linearly with the number of spins $N=L^d$,
where $d$ is the spatial dimension, we might expect from a simple random
walk argument that
$\tau \propto N^2$.
However, it was recently shown that for flat-histogram algorithms 
$\tau$ scales as
\begin{equation}
\tau \sim N^2L^{z},
\end{equation}
where the exponent $z$ is a measure of the deviation of the random walk from
unbiased Markovian behavior~\cite{tt}. In analogy to the dynamical critical
exponent for canonical simulations, the exponent $z$ quantifies the
critical slowing down for the simulated flat-histogram ensemble and the
chosen local update dynamics~\cite{EnsembleOptimization}. For the Ising
model on a square lattice it was shown in Ref.~\cite{tt} that $z \approx
0.74$.

In this paper, we determine the performance of the single-spin
flip Wang-Landau algorithm for the ferromagnetic Ising model as a function
of the dimension $d$. 
We find that the exponent $z$ is a decreasing function of $d$.
By using Monte Carlo simulations we determine that 
$z\approx 1.814$, 0.743, and 0.438 for $d=1$, 2, and 3, 
respectively (see Sec.~\ref{ttising}).
In the mean-field limit of infinite dimension we
find that $z=0$ using Monte 
Carlo simulations and analytic methods (Sec.~\ref{sec:mfim}). 
To round out our examination of the single-spin flip
algorithm, we also estimate $z$ for the $q=10$ and
$q=20$ Potts model in $d=2$
(see Sec.~\ref{sec:potts}).

We then discuss two complementary approaches that overcome the
critical slowing down of the single-spin flip flat-histogram algorithm. In
Sec.~\ref{cluster} we demonstrate that flat-histogram MC
simulations can be speeded up by {\em changing the spin dynamics}, that is,
by using cluster updates instead of local
updates~\cite{janke,kawashima}. 
Alternatively, we apply the recently developed ensemble optimization
technique~\cite{EnsembleOptimization} to {\em change the simulated 
statistical ensemble} and show that by sampling an optimized
histogram instead of a flat-histogram, the critical slowing down of the
single-spin flip simulation can also be eliminated (Sec.~\ref{optimized}). 
For the Ising model in one and two dimensions we find that 
both approaches result in optimal $N^2$-scaling of the random walk 
up to logarithmic corrections.

\section{Performance of flat-histogram algorithms with local updates}
\label{single}

In this section we study the performance of the single-spin flip
flat-histogram algorithm by measuring the mean first-passage time for the
$d=1$, 2, 3 and mean-field Ising models and the $q$-state Potts model in
$d=2$.

\subsection{The mean first-passage times for the Ising model
in $\mathbf{d=1}$, 2, and 3}\label{ttising}

\begin{figure}[t]
\includegraphics[scale=0.50]{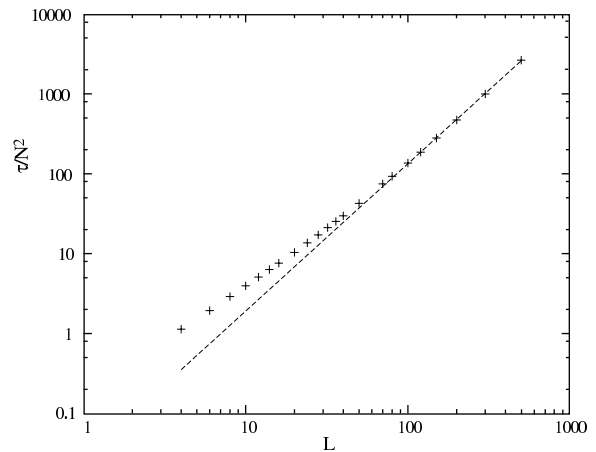}
\caption{\label{z-1D}
Scaling of the mean first-passage time in the energy interval $[-N,+N]$
for the flat-histogram random walker of the $d=1$ Ising model. 
The dashed line corresponds to a power-law fit in the range
$70 \leq N \leq 500$, which yields a critical exponent of 
$z = 1.814 \pm 0.014$.}
\end{figure}

For the $d=1$ Ising model, the mean first-passage time time was calculated
using the exact density of states and 50,000 MC measurements for each value of $N$.
A power-law fit in the range $70 \leq N \leq 500$ gives $z = 1.814 \pm 0.014$ 
as shown in Fig.~\ref{z-1D}. 
However, there is some curvature in the data which indicates a deviation 
from this power-law scaling and the effective value of $z$ seems to increase
as larger system sizes are taken into account. Thus we cannot eliminate the 
possibility that $z=2$.

A crude argument that $z \leq 2$ comes from considering the two lowest
states of a spin chain. The ground state has all spins aligned, and the
first excited state has a single misaligned domain bounded by two domain
walls. The distance between the domain walls is typically the order of
$L$, the linear dimension of the system. For single spin flip dynamics the
domain walls perform a random walk so that the typical number of sweeps
required to make the transition from the first excited state to the ground
state scales as $L^2$. If this diffusion time were to hold for all energies,
the result would be $z=2$. However, for higher energy states the domain 
wall diffusion time becomes smaller, so $z=2$ is an upper bound for the 
$d=1$ Ising model.

\begin{figure}[t]
\includegraphics[scale=0.5]{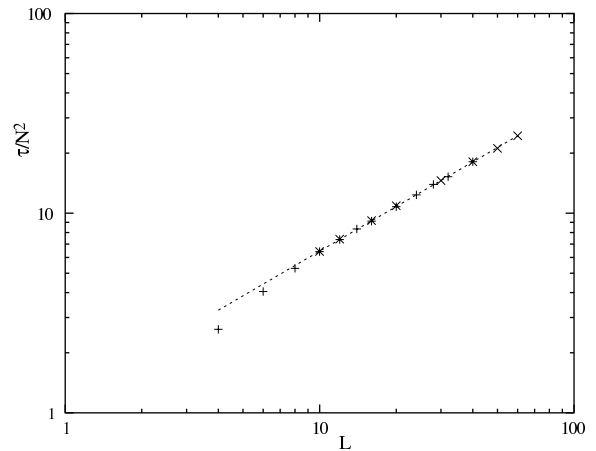}
\caption{\label{z-2D}Scaling of the mean first-passage time in the energy
interval $[-2N,+2N]$ for the flat-histogram random walker of the $d=2$ Ising
model. The dashed line corresponds to a power-law fit, which yields 
$z=0.743 \pm 0.002$.}
\end{figure}

The exact density states also was used for $d=2$~\cite{footnote1}. 
The combined results from Ref.~\cite{tt} and this work are shown in Fig.~\ref{z-2D}
where each data point represents a total of 50,000 measurements. 
A power-law fit of the critical exponent $z$ for $10 \leq L \leq 64$ gives
$z = 0.743\pm0.002$.

\begin{figure}[t]
\includegraphics[scale=0.50]{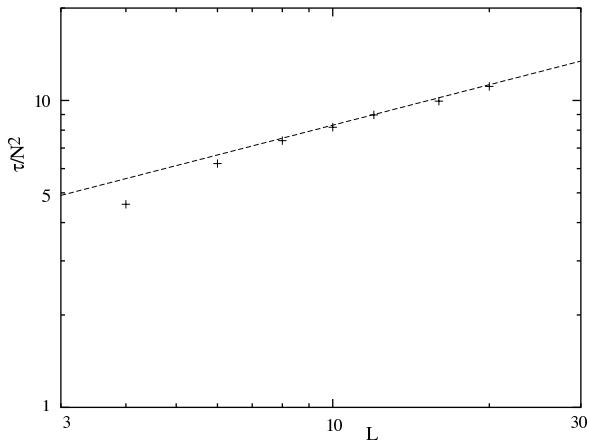}
\caption{\label{z-3D}Scaling of the mean first-passage time in the energy
interval $[-3N,+3N]$ for the flat-histogram random walker of the $d=3$ Ising
model. The dashed line corresponds to a power-law fit for $L \geq 8$, which
yields $z=0.743 \pm 0.002$.}
\end{figure}

For the $d=3$ Ising model the exact Ising density of states
cannot be calculated exactly, except for very small systems. We first used
the \wl\ algorithm~\cite{wl1,wl2} to estimate the density of states. The 
criteria for the completion of each iteration in the determination of $g(E)$
was that each energy be visited at least five times and that the variance
of the histogram be less than 10\% of the mean. Because $g(E)$ is symmetric
about $E = 0$, the calculated values for $g(E)$ and $g(-E)$ were averaged. 
For each value of $N$, $g(E)$ was independently calculated ten times and
then each $g(E)$ was used to obtain $\tau$ by averaging over 5,000 measurements. 
The results for the first-passage time $\tau$ were then averaged over ten independent runs. 
The results are shown in Fig.~\ref{z-3D} and yield
$z = 0.438 \pm 0.010$. 

\subsection{Mean-Field Ising model}\label{sec:mfim}
The results of Sec.~\ref{ttising} indicate that the value of $z$ decreases
with increasing dimension. This dependence suggests that it would be
interesting to compute $\tau$ for the Ising model in the mean-field limit.
We consider the ``infinite-range'' Ising model for which
every spin interacts with every other spin with an interaction strength
proportional to $1/N$. In this system the energy $E$ and magnetization
$M$ are simply related, and it is convenient to express the density of states
in terms of the latter. For each value of $N$ we did 50,000 MC measurements 
of $\tau$ in the range $4\le N \le 8000$.

We also calculated $\tau$ from a master equation. 
Because flipping a spin changes the magnetization
by $\pm 2$, the master 
equation takes the form (suitably modified near the extremes of $M = \pm N$)
\begin{align}
P(M,t+1) =& P(M,t) + T(M+2,M)\,P(M+2,t)\nonumber\\
&-[T(M,M+2) + T(M,M-2)]\,P(M,t)\nonumber\\
{}&+T(M-2,M)\,P(M-2,t),\label{prob1}
\end{align}
where $P(M,t)$ is the probability that the system has
magnetization $M$ at time $t$ and $T(M_1,M_2)$ is the probability of a
transition from a state with magnetization
$M_1$ to one with $M_2$. The transition probabilities $T$ are products of
the probability of choosing a spin in
the desired direction times the probability of 
accepting the flip. The latter is the
Wang-Landau probability, $W(M_1,M_2)$, which is given by
\begin{equation}
W(M_1,M_2) = \min\Big\lbrack 1,
{{g(M_1)}\over{g(M_2)}}\Big\rbrack.\label{WL}
\end{equation} 
The probability of choosing a spin in the desired direction
is determined as follows. In a transition in which the magnetization
increases, a down spin must be chosen. The probability of choosing a
down spin is the number of down spins, $N_\downarrow$, divided by the
total number of spins,
\begin{subequations}
\label{eq:nupdown}
\begin{equation}
{N_\downarrow(M) \over N} = {{N-M}\over{2\,N}}.
\end{equation}
Similarly, in a transition in which the magnetization
decreases, an up spin must be chosen. The probability is
\begin{equation}
{N_\uparrow(M)\over N} = {{N+M}\over{2\,N}}.
\end{equation}
\end{subequations}
The transition probabilities can now be written as:
\begin{subequations}
\label{eq:trans}
\begin{eqnarray}
T(M+2,M) &=& {N_{\uparrow}(M+2) \over N}W(M+2,M)\\
T(M,M+2) &=& {N_{\downarrow}(M) \over N}W(M,M+2)\\
T(M-2,M) &=& {N_{\downarrow}(M-2) \over N} W(M-2,M)\\
T(M,M-2) &=& {N_{\uparrow} (M) \over N}W(M,M-2)
\end{eqnarray}
\end{subequations}

The \wl\ probability $W(M_1,M_2)$ can be simplified for this
system in the following way: the density of states in terms of the 
magnetization is
\begin{equation}
g(M) = {N \choose N_\uparrow}.
\end{equation}
If $|M|$ increases in the
transition, that is, $|M_1| < |M_2|$, then $W(M_1,M_2) = 1$ because the
ratio of the density of states exceeds unity. If
$|M|$ decreases, there are two cases. For $M < 0$, $N_\uparrow$ increases by
$1$ and $M$ increases by 2, and we have
\begin{subequations}
\label{eq:w}
\begin{equation}
W(M,M+2) = {{g(M)}\over{g(M+2)}} = {{N \choose
N_\uparrow}\over{N\choose{N_\uparrow +1}}}
= {{N+M+2}\over{N-M}}.
\label{w+}
\end{equation}
If $M > 0$, $N_\uparrow$ decreases by 1 and $M$ decreases by
$2$, and
\begin{equation}
W(M,M-2) = {g(M) \over g(M-2)} = {{N\choose
N_\uparrow} \over {N \choose N_\uparrow -1} }
={N-M+2\over N+M}.
\label{w-}
\end{equation}
\end{subequations}
Equations~(\ref{WL}) and (\ref{eq:w}) can be
combined into the following simple expression
\begin{equation}
\label{eq:ws}
W(M,M\pm 2) = \min \Big[1,{{N \pm M + 2} \over N + |M|} \Big].
\end{equation}

\begin{figure}[t]
\includegraphics[scale=0.525]{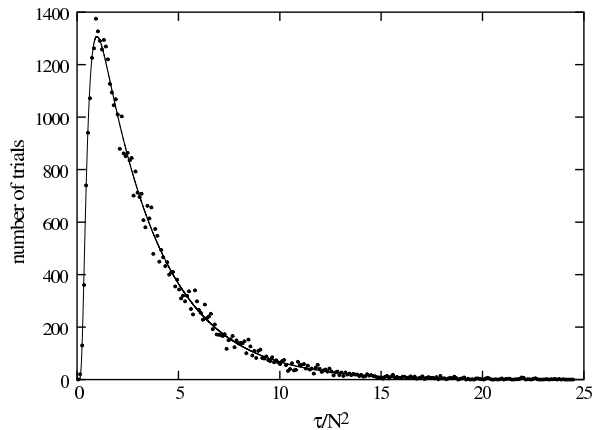}
\caption{\label{compare_master}Comparison of the distribution of
first-passage times for the mean-field Ising model computed by iterating the
master equation (line) and by direct simulation (points) for 
$N=32$. The master equation results were normalized to 50,000 MC estimates.} 
\end{figure}

\begin{figure}[t]
\includegraphics[scale=0.5]{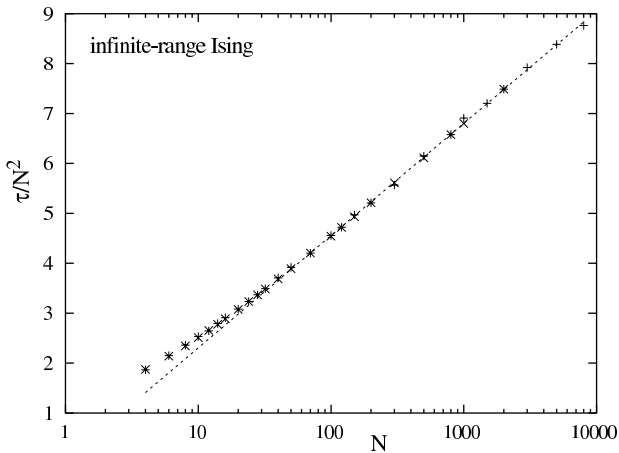}
\caption{\label{meanfield-z}
Scaling of the mean first-passage time in the magnetization interval $[-N,+N]$
for the flat-histogram random walk of the mean-field Ising model.
Note the linear scale on the $y$-axis.
The $+$ symbol denotes the simulation results and $\times$ denotes 
results derived from the master equation.
The dashed line corresponds to a logarithmic fit of the combined results for
$L \geq 50$. The slope is found to be $0.977 \pm 0.004$.}
\end{figure}

To compute the mean first-passage time, we take the state with magnetization
$N$ (the state with the highest magnetization) to be absorbing. 
The initial condition is
\begin{equation}
P(M,0) = \begin{cases}
1 & \text{$M = -N$} \\
0 & \text{otherwise} \;.
\end{cases}
\end{equation}
To compute $\tau$, we iterate Eq.~(\ref{prob1}) using
Eqs.~(\ref{eq:trans}) and (\ref{eq:w}) and compute
$\Delta P(N,t)$, the change in the value of $P(N,t)$ after each iteration
$t$. The mean first-passage time is then given by
\begin{equation}
\tau = \sum_t t\,\Delta P(N,t).
\end{equation}
A comparison of the distribution of first-passage times calculated from
the master equation and from a direct simulation is shown in
Fig.~\ref{compare_master}. 
As shown in Fig.~\ref{meanfield-z}, the scaling of the mean first-passage 
time with $N$ is consistent with $\tau/N^2 \sim \ln N$. 
The fit includes both the master equation and simulation results.

The logarithmic behavior of $\tau$ can be determined analytically from the
transition rates using the first-passage time methods described in
Ref.~\cite{Redner}. Let $t(m)$ be the mean first-passage time starting from
magnetization $m=M/N$ and ending at the absorbing boundary $m=1$ with a
reflecting boundary at $m=-1$. We seek $\tau=t(-1)$. The mean
first-passage time satisfies,
\begin{eqnarray}
\label{eq:rec}
t(m) &= & [t(m-\delta m) + \delta t] T_{-}(m) \nonumber\\
&&{} + [t(m+\delta m) + \delta
t]T_{+}(m) \nonumber \\
 &&{}+ [t(m) + \delta t] T_0(m),
\end{eqnarray}
where $T_{-}(m) = T(M,M-2)$, $T_{+}(m) = T(M,M+2)$ are the transition
probabilities, $T_0 = 1 - T_{-} - T_{+}$ is the waiting probability, $\delta
m = 2/N$ is the magnetization step size, and $\delta t =1$ is the time unit
for a step. Equation~(\ref{eq:rec}) can be cast in differential form by
expanding
$t(m)$ to second order in $\delta m$,
\begin{equation}
\label{eq:diff} 1 + h(m) t^\prime(m) + f(m) t^{\prime\prime}(m) = 0,
\end{equation}
where
\begin{equation}
\label{eq:f}
f(m)= (\delta m^2/2 \delta t)[T_{+}(m)+T_{-}(m)],
\end{equation}
and
$h(m)=(\delta m/\delta t) [T_{+}(m)-T_{-}(m)]$. The absorbing boundary
requires that
\begin{equation}
\label{eq:b1} t(1) = 0.
\end{equation}
At the reflecting boundary, the transition to the left vanishes,
$T_{-}(-1)=0$, and we have
\begin{eqnarray}
\label{expand}
t(-1) &=& [t(-1+\delta m) + \delta t] T_{+}(-1) \nonumber \\
&&{}+ [t(-1) + \delta t]
[1-T_{+}(-1)].
\end{eqnarray}
If we expand Eq.~(\ref{expand}) to first order in $\delta m$, we obtain the
boundary condition
\begin{equation}
t^\prime(-1) = - \frac{\delta m}{2f(-1)},
\end{equation}
where $f(-1)= (\delta m^2/2 \delta t) T_{+}(m)$

The flat-histogram feature of the \wl\ method means that the random walk
satisfies the simple detailed balance condition,
\begin{subequations}
\begin{eqnarray}
T_{+}(m)&=&T_{-}(m+\delta m) \\
T_{-}(m)&=&T_{+}(m-\delta m).
\end{eqnarray}
\end{subequations}
We expand the detailed balance conditions to first order in $\delta m$
and find a relation between the coefficients in Eq.~(\ref{eq:diff}),
\begin{equation}
h(m)=f^\prime(m),
\end{equation}
with the result that Eq.~(\ref{eq:diff}) reduces to 
\begin{equation}
\label{eq:diff1} 1 + f^\prime(m) t^\prime(m) + f(m) t^{\prime\prime}(m) = 0.
\end{equation}
If we take into account the reflecting boundary condition, the solution to
the differential equation is 
\begin{equation}
t(m)= \tau -\!\int_{-1}^{m} \frac{x+1 + \delta m/2}{f(x)} dx,
\end{equation}
where $\tau=t(-1)$ is the first-passage time we are seeking. We use the
absorbing boundary condition to find that
\begin{equation}
\tau = \!\int_{-1}^{1} \frac{x+1 + \delta m/2}{f(x)} dx.
\end{equation}
Because $f(m)$ is an even function, the first term in the integrand
vanishes. We substitute $\delta m= 2/N$ and use
Eq.~(\ref{eq:f}) for $f(m)$ to
obtain the leading behavior of $\tau$,
\begin{equation}
\tau= 2 N^2 \!\int_{-1}^{1} \frac{1 }{1- T_{0}(x)} dx.
\end{equation}
{}From Eqs.~(\ref{eq:nupdown}), (\ref{eq:trans}), and (\ref{eq:ws}) we have
that $T_{0}(m)= |m|- 1/N$, which yields the desired result
\begin{equation}
\label{eq:dttimeO}
\tau = 2 N^2 \!\int_{-1}^1 \frac{d m}{1-|m| + 1/N} \sim N^2\log N.
\end{equation} Note that the integral can be interpreted as the average
waiting time and that this average is dominated by the two boundaries.

\subsection{The Potts model}\label{sec:potts}

In Ref.~\cite{EnsembleOptimization} it was shown that the non-vanishing
scaling exponent $z$ quantifies the critical slowing down of the
flat-histogram simulation near a critical point. Most strikingly, this
slowing down can be observed as a suppression of the local diffusivity of
the flat-histogram random walker (see Sec.~\ref{optimized}). The Ising
model undergoes a second order phase transition. Here we extend our
analysis to first-order phase transitions as is observed in
$q$-state Potts models with $q>4$ on a square lattice.

\begin{figure}[t]
\includegraphics[scale=0.52]{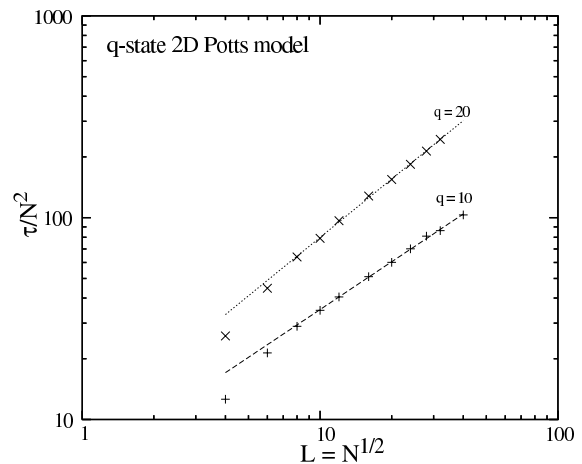}
\caption{\label{zPotts}
Scaling of the mean first-passage time in the energy interval $[-2N,0]$
for the flat-histogram random walk of the $d=2$, $q=10$ and $q=20$ Potts
models. The dashed lines correspond to power-law fits for $L \geq 8$ and
yield $z=0.786 \pm 0.010$ and $z=0.961 \pm 0.008$ for
the $q=10$ and
$q=20$ Potts models, respectively.}
\end{figure}

Because the exact density of states is not known for the $q$-state Potts
model, we follow the same procedure as for the Ising model in $d=3$ making
use of the
\wl\ algorithm. Based on these estimates for the density of states, we 
measured the mean first-passage time as a function of the linear system size
$L$ as is illustrated in Fig.~\ref{zPotts}. For $q=10$ we find that the
critical exponent $z$ becomes 
$z = 0.786 \pm 0.010$, which is close to our result for the $d=2$ Ising model
thereby showing the close proximity of this weak first-order phase
transition to a continuous phase transition. For the $q=20$ Potts model we
find $z=0.961 \pm 0.008$, illustrating increasing critical slowing down as
the strength of the first-order transition increases. However, we cannot
rule out exponential slowing down for even larger values of $q$ as was
observed for multicanonical simulations of droplet condensation in the $d=2$
Ising model~\cite{Neuhaus:2003}.

\section{Changing the dynamics:\newline Cluster Updates}\label{cluster}
In Sec.~\ref{single} we found that single-spin flip flat-histogram Monte
Carlo simulations suffer from a critical slowing down, that is, 
$z>0$, for the Ising model in $d=1$, 2, and 3. The critical
slowing down associated with single-spin flip algorithms near critical
points in the canonical ensemble has been reduced by the introduction of
cluster algorithms such as the Swendsen-Wang algorithm~\cite{sw} and the
Wolff algorithm~\cite{wolff}. In this section we follow a similar approach
and combine efficient cluster updates with flat-histogram simulations.

In conventional cluster algorithms~\cite{sw,wolff} simulating the canonical
ensemble, clusters of parallel spins are built up by adding aligned spins to
the cluster with a probability $p(\beta)$, which
explicitly depends on the temperature $\beta = 1/T$. When simulating a
broad-histogram ensemble in energy there is no explicit notion of 
temperature, and thus no straightforward analogue of a cluster
algorithm. Recently, Reynal and Diep suggested a solution of this problem
by using an estimate of the microcanonical temperature $\beta(E) =
dS(E)/dE$~\cite{diep}. 

A complementary approach is to sample a broad-histogram ensemble using an
alternative representation of the system's partition function for which it
is possible to genuinely introduce cluster updates. An example is the
multibondic method introduced by Janke and Kappler~\cite{janke} which uses
the Fortuin-Kasteleyn (FK) representation~\cite{fk} in the context of
multicanonical sampling. Although the multibondic method performs {\em
local} updates of the graph in the FK representation and cluster updates of
the spin configurations, Yamaguchi and Kawashima were able to show that a
\textit{global} and \textit{rejection free} update of the graph in the FK
representation is possible~\cite{kawashima}. We will discuss methods based
on graph representations in Sec.~\ref{graph}.

The above methods can only be applied to Ising and Potts models.
In Sec.~\ref{multigraph} we introduce a new representation using
multigraphs that allows cluster updates for continuous spin models in the
context of broad-histogram ensembles.

In the following we assume that the Hamiltonian of the system has the form 
\begin{equation}
H[\sigma] = - \sum_{\langle ij \rangle} h_{ij}[\sigma_{i},\sigma_{j}] \;,
\label{hamdef}
\end{equation}
where $\sigma$ denotes the configuration of the system, $\sigma_i$
the spin of site $i$, and the sum is over all
bonds $\langle ij \rangle$ that define the lattice of the system.

\subsection{Graph representation}\label{graph}
We first review the cluster methods based on an additional graph variable and 
present results showing that $z=0$ for multibondic flat-histogram
simulations of Ising models. 
In the spin representation the canonical partition function of the
Ising-Potts model is given by
$Z=\sum_{\sigma} W_{\sigma}(\sigma)$, where the summation is over all spin
configurations, and the weight of a spin configuration is
$W_\sigma(\sigma)=e^{-\beta H[\sigma]}$. 

The $q$-state Ising-Potts system also can be represented by a sum over
all graphs $\omega$ that can be embedded into the lattice. The 
partition function is then
\begin{equation}
\label{eq:bondZ} Z=\sum_{\omega}{W_{\omega}(\omega)},
\end{equation}
and the weight of a graph $\omega$ is given by
\begin{equation}
\label{eq:fkweight}
W_\omega(\omega) = p^{n(\omega)}(1-p)^{n_b - n(\omega)}q^{c(\omega)}.
\end{equation}
$n(\omega)$ is the number of of bonds in the graph $\omega$,
$c(\omega)$ the number of connected components (counting isolated sites), 
and $n_b$ is the total number of bonds in the lattice.
Because the graph $\omega$ can be viewed as a subgraph of the lattice, its
bonds are sometimes referred to as ``occupied bonds,'' and the bond
probability of the ferromagnetic Ising-Potts model with 
$h_{ij}[\sigma_i,\sigma_j] = \delta_{\sigma_i,\sigma_j}$
is given by~\cite{fk}
\begin{equation}
\label{eq:p}
p(\beta)= 1-e^{-\beta}.
\end{equation}

A third representation of the Ising-Potts model is the spin-bond
representation, which is employed in the Swendsen-Wang algorithm~\cite{sw}.
In the spin-bond representation the system is characterized by both spins
and bonds, with the requirement that a bond can be occupied only if it
is satisfied, that is if the two spins connected by a bond in $\omega$
have the same value. In this representation the partition function is
\begin{equation}
\label{eq:spinbondZ} 
Z = \sum_{\sigma,\omega}{W_{\sigma \omega}(\sigma,\omega)},
\end{equation}
and the weight is given by
\begin{equation}
W_{\sigma \omega}(\sigma,\omega) = p^{n(\omega)}(1-p)^{n_b - 
n(\omega)}\Delta(\sigma,\omega),
\end {equation}
where $\Delta(\sigma,\omega)=1$ if all occupied bonds are satisfied
and zero otherwise.

It is natural to introduce a density of states for each of the three
representations:
For the spin representation in terms of the energy $E$, we write
\begin{equation}
g_\sigma(E) = \sum_{\sigma} \delta({H[\sigma]}-E).
\end{equation}
For the bond representation in terms of the number of occupied bonds 
$n$ and number of clusters $c$, we have
\begin{equation}
g_\omega(n,c) = \sum_{\omega} \delta({n(\omega)}-n) \, \delta(c(\omega)-c).
\end{equation}
And for the spin-bond representation in terms of the number of occupied 
bonds $n$, we write
\begin{equation}
g_{\sigma\omega}(n) = \sum_{\omega,\sigma} \delta({n(\omega)}-n) 
\, \Delta(\sigma,\omega).
\end{equation}
The corresponding forms of 
the partition function are
\begin{align}
Z &= \sum_E{g_\sigma(E) \, e^{-\beta E}} & \mbox{(spin)} \label{spinrep}\\
Z &= \sum_{n,c}{g_\omega(n,c) \, p^{n} \, (1-p)^{n_b - n} \, q^{c}} 
&\mbox{(bond)}\\
Z &= \sum_{n} g_{\sigma\omega}(n) \, p^{n} \, (1-p)^{n_b - n}. 
& \mbox{(spin-bond)}
\end{align}
The \wl\ algorithm can be applied in
all three representations.
As discussed in Sec.~\ref{single}, the spin
representation has a relatively large value of $z$.

\begin{figure}
\includegraphics[scale=0.8]{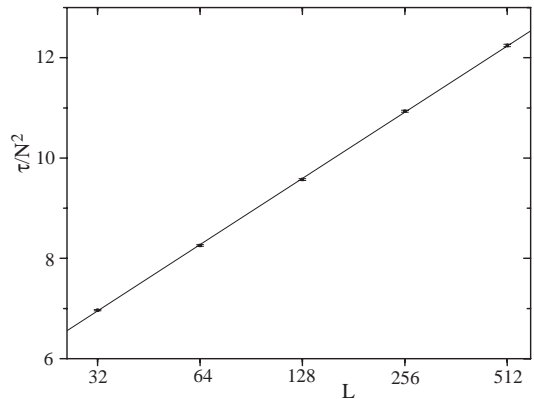}
\caption{
Scaling of the mean first-passage time applying cluster updates in the bond 
representation for the $d=1$ Ising model. Note the linear scale of the
$y$-axis.}

\label{fig:FKWL}
\end{figure}

The bond representation generally requires a two-dimensional histogram, but
for the $d=1$ Ising-Potts model, the number of clusters $c$ is completely
determined by the number of occupied bonds $n$ and a one-dimensional
histogram is sufficient. We simulated the $d=1$ Ising model in the bond
representation and found that the mean first-passage time scales as $N^2\log
L$, as shown in Fig.~\ref{fig:FKWL}. It is noteworthy that the domain wall
arguments used to explain the large value of $z$ in one dimension do not
apply in the bond representation. 

In higher dimensions $n$ does not determine $c$, and it is simpler to use
the spin-bond representation. The resulting multibondic
algorithm~\cite{kawashima} is very similar to the Swendsen-Wang
algorithm, and can be summarized as follows,

\begin{enumerate}
\item Choose a bond at random.

\item If the bond is satisfied and occupied (unoccupied), make it 
unoccupied (occupied) with probability
$\min[ g_{\omega\sigma}(n)/g_{\omega\sigma}(n^\prime),1]$, 
where $n$ and $n^\prime$ are the number
of occupied bonds before and after the change, respectively. Update the
observables.

\item If the bond is unsatisfied, the system stays in its original 
state and the observables are updated.

\item After one sweep of the lattice, identify clusters (connected components) and assign spin values to them with equal probability.

\end{enumerate}

The algorithm is ergodic and satisfies detailed balance. The
ergodicity is obvious from the fact that the Swendsen-Wang algorithm is
ergodic. In general, the detailed balance relation can be written as,
\begin{align}
\label{eq:balance} 
&\pi(\sigma,\omega)p(\sigma\rightarrow\sigma^\prime\vert\omega)
p(\omega\rightarrow\omega^\prime\vert\sigma^\prime) \nonumber \\
&=\pi(\sigma^\prime,\omega^\prime)p(\omega^\prime\rightarrow\omega
\vert\sigma^\prime)p(\sigma^\prime\rightarrow\sigma\vert\omega),
\end{align}
where $\pi(\sigma,\omega)$ is the equilibrium probability of 
microstate $(\sigma,\omega)$, and
$p(\sigma\rightarrow\sigma^\prime\vert\omega)$ is the transition 
probability from $\sigma$ to $\sigma^\prime$
given that the bond configuration is fixed at $\omega$.

The algorithm is designed to sample the distribution,
\begin{equation}
\pi(\sigma,\omega)=\frac{1}{(n_b+1)g_{\sigma\omega}({n(\omega)})}\Delta 
(\sigma,\omega).
\end{equation}
The transition probability of a spin flip is
\begin{equation}
p(\sigma\rightarrow\sigma^\prime\vert\omega)=\frac{1}{q^{c(\omega)}}
\Delta(\sigma,\omega),
\end{equation}
and the transition probability of a bond change 
($\omega^\prime\not=\omega$) is given by,
\begin{equation}
p(\omega\rightarrow\omega^\prime\vert\sigma)=\frac{1}{{n_b}}
\min\Big[\frac{g_{\sigma\omega}({n(\omega)})}{g_{\sigma\omega}
({n(\omega^\prime)})},1\Big],
\end{equation}
because the probability of choosing a bond to change is 
$1/{n_b}$, and the
probability
of accepting the change is
$\min\big[{g_{\sigma\omega}({n(\omega)})}/ 
{g_{\sigma\omega}({n(\omega^\prime)})},1\big]$. It is easy to verify
that the detailed balance relation, Eq.~(\ref{eq:balance}), is satisfied.

\begin{figure}[t]
\includegraphics[scale=0.8]{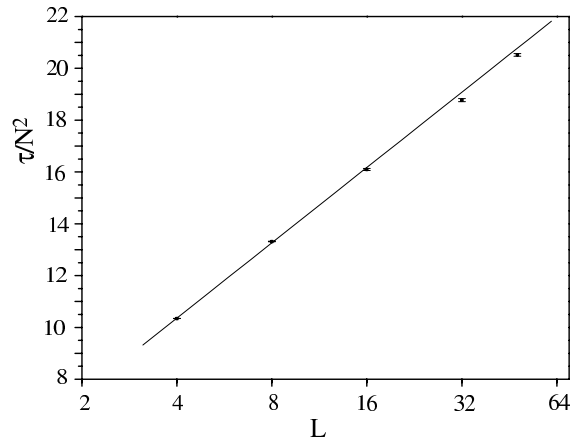}
\caption{Scaling of the mean first-passage time using cluster updates in
the spin-bond representation of the $d=2$ Ising model. Note the linear
scale of the $y$-axis.}
\label{fig:WL2D}
\end{figure}

We have applied this algorithm to the $d=2$ Ising model and measured the
mean first-passage time, which also scales as $N^2\log L$, as shown in
Fig.~\ref{fig:WL2D}. At least 10$^5$ first-passage
times were measured for each value of $L$.

\subsection{Spin-multigraph representation}\label{multigraph}
We now consider a representation of the partition function that can be
used to implement cluster updates in flat-histogram simulations of
\textit{continuous} spin models such as $O(n)$ models.
Like the spin-bond representation it has a configuration and a 
graph variable.
In contrast to the spin-bond representation, the graph is a multigraph
and can therefore have multiple bonds between two vertices.
Before introducing this representation, we briefly review
how a simple high-temperature series representation can be
sampled in a flat-histogram simulation.

We write the partition function as a series in the inverse
temperature $\beta$
\begin{equation}
Z = \sum_{n} g_{\beta}(n) \beta^{n},
\label{htpf}
\end{equation}
with a corresponding density of states given by
\begin{equation}
g_{\beta}(n) = \sum_{\sigma} \frac{(-H[\sigma])^{n}}{n!} \label{gbeta}. 
\end{equation}
If $H[\sigma] \le 0$, then the weight of a configuration-order pair 
\begin{equation}
W_\beta(\sigma,n) = \frac{(-H[\sigma])^{n}}{n!},
\label{def:weightbeta}
\end{equation}
is always positive, and we can perform a flat-histogram simulation in the
extended phase space $(\sigma,n)$ using $\sigma$ and $n$ updates.
If the $\sigma$ and $n$ updates are performed independently of each
other, an $n$ update 
$n \rightarrow n'$ (for example $n' = n \pm 1$) is accepted
with probability
\begin{equation}
P_{n \to n'} = 
\min \!\Big[ 
(-H[\sigma])^{n'-n} \frac{n!}{n'!}
\frac{g_{\beta}(n)}{g_{\beta}(n')}, 1 \Big].
\label{up_n}
\end{equation}
A configuration update 
$\sigma \rightarrow \sigma'$ (for example the change of the configuration of
a single site $\sigma_i \rightarrow \sigma'_i$) at fixed $n$ is accepted with
probability
\begin{equation}
P_{\sigma \to \sigma'} = 
\min \!\Big[ \Big( \frac{H[\sigma']}{H[\sigma]} \Big)^{n}, 1 \Big].
\label{up_simple}
\end{equation}
In a practical computation $n$ is truncated at some cutoff $\Lambda$
restricting the largest inverse temperature that can be reached to 
$\beta_{\max} \sim \Lambda/V$~\cite{troyer}, where $V$ is the volume
of the system. 

Because the weight of a configuration as defined in 
Eq.~(\ref{def:weightbeta}) is not a product of bond
weights on the lattice, the $(\sigma,n)$-representation
cannot be directly used to implement cluster updates.
To obtain such a representation we start from
\begin{equation}
e^{-\beta H[\sigma]} = \prod_{\langle ij \rangle} e^{\beta
h_{ij}[\sigma_i,\sigma_j]},
\end{equation}
and expand each exponential term in $\beta$ with a separate
integer variable $\omega_{ij}$
\begin{equation}
e^{\beta h_{ij}[\sigma_i,\sigma_j]} = \sum_{\omega_{ij}=0}^{\infty} 
\frac{h_{ij}[\sigma_i,\sigma_j]^{\omega_{ij}}
\beta^{\omega_{ij}}}{\omega_{ij}!}.
\end{equation}
We identify each set of $\omega_{ij}$ on the lattice with a multigraph
$\omega$ that has $\omega_{ij}$ bonds between site $i$
and site $j$. 
If we write $n(\omega) = \sum \omega_{ij}$ for the total
number of bonds in the multigraph $\omega$, $g_\beta(n)$ as defined
in Eq.~(\ref{gbeta}) is given by
\begin{equation}
g_{\beta}(n) = \sum_{\sigma,\omega} \delta(n(\omega)-n)
\prod_{\langle ij \rangle}
\frac{h_{ij}[\sigma_i,\sigma_j]^{\omega_{ij}}}{\omega_{ij}!},
\label{pf_multi}
\end{equation}
where the sum is over all spin configurations $\sigma$ and all
multigraphs $\omega$ on the lattice.
The corresponding partition function is given by Eq.~(\ref{htpf}), and
the weight of a configuration-multigraph pair is
given by
\begin{equation}
W_{\beta}(\sigma,\omega)
= \prod_{\langle ij \rangle} W_{ij}(\sigma_i,\sigma_j,\omega_{ij}) ,
\label{wsigomega}
\end{equation}
where
$W_{ij}(\sigma_{i},\sigma_{j},\omega_{ij}) = {h_{ij}[\sigma_i,\sigma_j]^{\omega_{ij}}}/{\omega_{ij}!}$
is the weight of the bond $\langle ij \rangle$ in the lattice.
The energies in the system 
have to be shifted so that $h_{ij}[\sigma_i,\sigma_j] \ge 0$ to 
ensure that all weights are positive.
Again two types of updates, one in the multigraph $\omega$ and one in
the configuration $\sigma$ are needed. Before discussing
them in detail, we mention some differences of the 
spin-multigraph representation compared to the
representations used in Sec.~\ref{graph}.
In the spin-multigraph representation we cannot in general integrate 
explicitly over all configurations to obtain a multigraph-only 
representation similar to the bond representation 
except for Ising or Potts models.
It also is not possible to apply the global graph update of
Yamaguchi and Kawashima~\cite{kawashima} to the multigraph
so that local updates have to be used.
Finally the representation can be viewed as the classical limit
of the stochastic series expansion method (SSE) for quantum systems~\cite{sse}.
The order of the operators in the operator-string used in the SSE
is unimportant for a classical system, so that one only has to
count how often an operator $h_{ij}[\sigma_i,\sigma_j]$ occurs ($\omega_{ij}$).

If we use a flat-histogram algorithm in the order $n$, a multigraph update 
$\omega \rightarrow \omega'$ 
(as before $\omega'_{ij} = \omega_{ij} \pm 1$ for one randomly chosen bond 
$\langle ij \rangle$ is a good choice) will be accepted with probability
\begin{equation}
P_{\omega \to \omega'} = 
\min \Big[ 
\frac{g_{\beta}(n(\omega))}{g_{\beta}(n(\omega'))}
\prod_{\langle ij \rangle} 
\frac{W_{ij}(\sigma_{i},\sigma_{j},\omega'_{ij})}
{W_{ij}(\sigma_{i},\sigma_{j},\omega_{ij})}, 1 \Big].
\label{up_nij}
\end{equation}

\begin{figure}[t]
\includegraphics[scale=0.325]{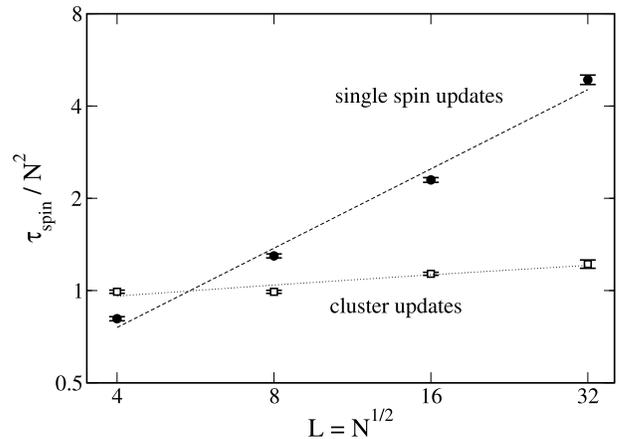}
\caption{\label{figure_ising}Scaling of the mean first-passage time for the
$d=2$ Ising model using the simple high temperature representation
with single spin updates ($\bullet$) and using the spin-multigraph
representation with Swendsen-Wang cluster updates ($\square$). 
The lines are drawn as a guide to the eye.}
\end{figure}

As the weight of a spin-multigraph configuration is a product of 
bond weights, a cluster update scheme
such as the Swendsen-Wang~\cite{sw} or Wolff algorithm~\cite{wolff} 
can be used to update the spin configuration.
After choosing a flip operation for the spins,
the only difference to the canonical cluster algorithms is
the probability of adding a site to a Swendsen-Wang cluster: 
If $W_{ij}$ is the weight if either none or both sites are flipped 
and $\tilde{W}_{ij}$ the weight after a flip of one of the sites
(note that this requires 
$W_{ij}(\sigma_{i},\sigma_{j}) = W_{ij}(\sigma'_{i},\sigma'_{j})$ 
and $W_{ij}(\sigma'_{i},\sigma_{j}) = W_{ij}(\sigma_{i},\sigma'_{j})$),
the probability of adding a site to a cluster is given by
\begin{equation}
\label{eq:thisP}
P_{\mathrm{connect}}=1-\min\Big[\frac{\tilde{W}_{ij}}{W_{ij}}, 1 \Big].
\end{equation}
For the canonical ensemble simulation of the Ising model with
$W(\uparrow,\uparrow,\beta) = e^{\beta}$ and
$W(\uparrow,\downarrow,\beta) = 1$, Eq.~(\ref{eq:thisP})
reduces to the well known $1-e^{-\beta}$ 
for parallel spins as discussed before. 
In the spin-multigraph representation the probability to add
a site to a cluster is
\begin{equation}
P_{\mathrm{connect}}=1-\min
\Big[\Big(\frac{h_{ij}[\sigma_i,\sigma'_j]}{h_{ij}[\sigma_i,\sigma_j]}\Big)^{\omega_{ij}},
1 \Big].
\label{up_cluster}
\end{equation}
Thus only sites that are connected by at least one bond of the multigraph
can be added to the cluster.
For the Ising model with $h_{ij}[\sigma_i,\sigma_j] =
\delta_{\sigma_i,\sigma_j}$, the weight of a configuration $(\sigma,\omega)$
as defined in Eq.~(\ref{wsigomega}) is
\begin{equation}
W_{\beta}(\sigma,\omega) = \Delta(\sigma,\omega)
\frac{\beta^{n(\omega)}}{\prod_{\langle ij \rangle}\omega_{ij}!},
\end{equation}
where $\Delta(\sigma,\omega) = 1$ if all bonds in the multigraph
$\omega$ are satisfied and $0$ otherwise, and
$P_\mathrm{connect}$ simplifies to $1$ if two sites are connected 
by one or more bonds in the multigraph and to $0$ if they are not.
$O(n)$ models can be simulated by performing the spin update with respect to
a mirror plane that is randomly chosen for one cluster update~\cite{wolff}.
\begin{figure}[t]
\includegraphics[scale=0.325]{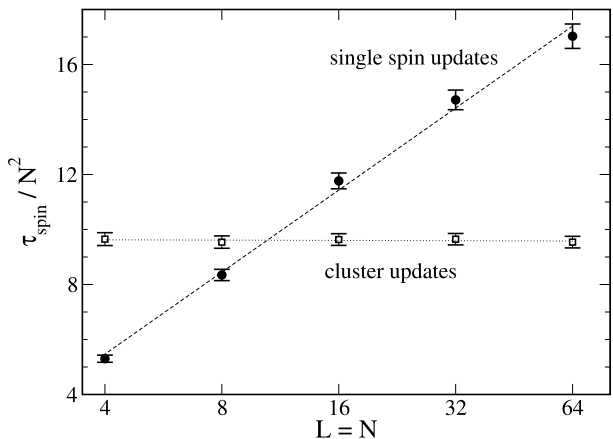}
\caption{\label{figure_xy}
Scaling of the mean first-passage time to reach a low energy state for the
$d=1$ XY model using single spin updates ($\bullet$) and Swendsen-Wang
cluster updates ($\square$) in the spin-multigraph. Lines are drawn as a
guide to the eye.}
\end{figure}

In a local update scheme, a configuration update 
$\sigma \rightarrow \sigma'$ at fixed $\omega$ will be accepted with
probability
\begin{equation}
P_{\sigma \to \sigma'} = 
\min \!\Big[ \prod_{\langle ij \rangle} \Big( 
\frac{h_{ij}[\sigma'_i,\sigma'_j]}{h_{ij}[\sigma_i,\sigma_j]}
\Big)^{\omega_{ij}}, 1 \Big].
\label{up_simpleij}
\end{equation}

We have applied a flat-histogram sampling of this representation to the
$d=2$ Ising model with periodic boundary conditions 
and to the $d=1$ XY model with open boundary conditions.
To compare the effectiveness of the cluster updates to the single-spin flip
updates, we measure the mean first-passage time $\tau_{\mathrm{spin}}$
from a random state $\sigma$ at $n=0$, which is equivalent to $\omega_{ij}=0$
for all $\langle ij \rangle$, to an ordered state.
For the Ising model this state is chosen to be the ground state and
for the $d=1$ model any state which is within 2.5\% of the width 
of the spectrum to the ground state. 
We perform a fixed ratio of spin to multigraph updates and measure the
mean first-passage time in spin updates for different ratios of spin to
multigraph updates.
$\tau_\mathrm{spin}$ converges to a fixed number 
with an increasing fraction of multigraph updates, and 
for sufficiently many multigraph updates we can assume the multigraph 
is equilibrated for a particular configuration.
The cutoff $\Lambda$ is chosen so that the average $n$ for which the ground 
state is reached is much smaller than $\Lambda$. 

Figure~\ref{figure_ising} shows the scaling of
$\tau_\mathrm{spin}$ for the 
$d=2$ Ising model for single-spin flips in the simple representation
of Eq.~(\ref{def:weightbeta}) and for Swendsen-Wang cluster updates in the
spin-multigraph representation.
The single-spin flip updates show a power law behavior with 
$z = 0.85 \pm 0.06$, while the cluster updates can be fitted by
a logarithmic behavior or a power law with $z = 0.10 \pm 0.02$.
Figure~\ref{figure_xy} shows the scaling of $\tau_\mathrm{spin}$ 
for the $d=1$ XY model using single spin updates and 
Swendsen-Wang cluster updates in the spin-multigraph
representation.

\section{Changing the ensemble: Optimizing the sampled histogram}
\label{optimized}

An alternative to changing the dynamics from local to cluster updates for
flat-histogram sampling is to optimize the simulated statistical ensemble 
and retain local spin-flip updates. 
When simulating a flat-histogram ensemble, the random walker is slowed
down close to a critical point, which is reflected in the suppressed
local diffusivity. To overcome this critical slowing down, we can use this
information and define a new statistical ensemble $w(E)$ by feeding back the
local diffusivity $D(E)$~\cite{EnsembleOptimization}. After the feedback
additional weight is shifted toward the critical point. The new histogram is
no longer flat, but exhibits a peak at the critical energy, that is,
resources in the form of local updates are shifted toward the critical
point. Ultimately, this feedback procedure results in an optimal histogram
$n_w(E)$ that is proportional to the inverse of the square root of the local 
diffusivity $D(E)$
\begin{equation}
n_w(E) \propto \frac{1}{\sqrt{D(E)}}.
\end{equation}

\begin{figure}
\includegraphics[scale=0.325]{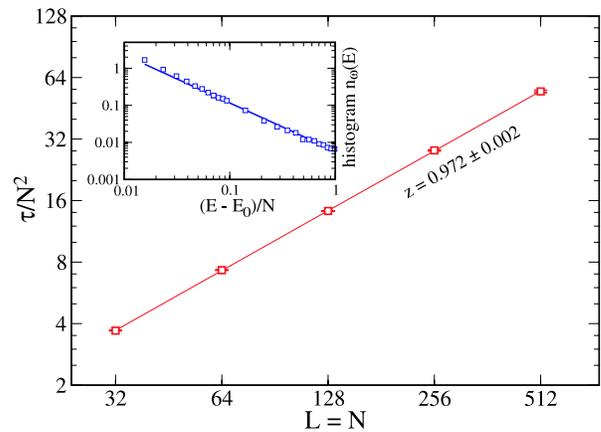}
\caption{Scaling of the mean first-passage time of the optimized ensemble
for the $d=1$ Ising model using Metropolis updates. The
solid line corresponds to a power-law fit with 
exponent $0.972 \pm 0.002$. 
The inset shows the optimized histogram which exhibits a power-law divergence 
$[(E-E_0)/N]^{1.30}$.}
\label{fig:OptEnsembleMetropolis}
\end{figure}

For the $d=2$ Ising model it was shown in
Ref.~\onlinecite{EnsembleOptimization} that the mean first-passage time
scales as $\tau \propto (N \ln N)^2$ for the optimized ensemble and satisfies
the scaling of an unbiased Markovian random walk up to logarithmic
corrections. It also was found that the distribution of statistical errors in
the calculated density of states $g(E) = n_w(E)/w(E)$ is
uniformly distributed in energy, in contrast to calculations based on
flat-histogram sampling.

Here we show results for the optimized ensemble of the $d=1$ 
Ising model for both Metropolis and $N$-fold way
updates~\cite{NFoldWay}. For single-spin flip Metropolis updates, we find
that the histogram is shifted toward the ground-state energy $E_0 = -N$ and
follows a power-law divergence,
$n_w(E)
\propto [(E-E_0)/N]^{1.30 \pm 0.01}$ (see the inset of 
Fig.~\ref{fig:OptEnsembleMetropolis}). However, the mean first-passage time
of the random walk in the energy interval $[-N, 0]$ still exhibits a
power-law slowdown with
$z=0.972 \pm 0.002$ as illustrated in the main panel of Fig.~\ref{fig:OptEnsembleMetropolis}.
This remaining slowdown may originate from the slow dynamics that occurs
whenever two domain walls reside on neighboring bonds, and a spin flip of
the intermediate spin is suggested with a probability of $1/N$. This
argument suggests that changing the dynamics from simple Metropolis updates
to $N$-fold way updates would strongly increase the probability of
annihilating two neighboring domain walls. 
		
\begin{figure}[t]
\includegraphics[scale=0.325]{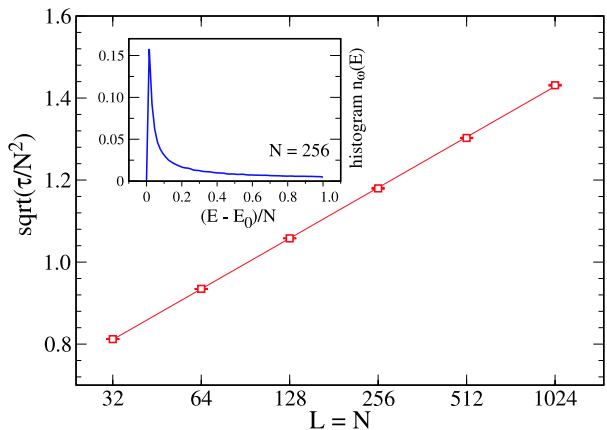}
\caption{Scaling of the mean first-passage time of the optimized ensemble
for the 1D Ising model using $N$-fold way updates (note the linear
scale on the $y$-axis). The solid line corresponds to a logarithmic fit. 
The inset shows the optimized histogram.}
\label{fig:OptEnsembleNfoldWay}
\end{figure}

We optimized the ensemble for $N$-fold way updates and found that the mean
first-passage time is further reduced and scales as $\tau \propto (N
\ln N)^2$ as shown in Fig.~\ref{fig:OptEnsembleNfoldWay}. This scaling
behavior corresponds to the results for the
optimized ensemble for the $d=2$ Ising model~\cite{EnsembleOptimization}.
However, for the $d=2$ model the optimized ensembles for both Metropolis and
$N$-fold way updates resulted in the same scaling behavior of the form $\tau
\propto (N \ln N)^2$.

\section{Conclusions}

We have studied the performance of flat-histogram simulations
of Ising and Potts models in different dimensions.
For one, two, and three dimensions the simulations show critical slowing
down with a critical exponent $z>0$. This behavior is analogous to the
critical slowing down of canonical ensemble simulations of these models using
single spin updates. 
Our numerical results show that $z$ decreases as a function of the
dimensionality $d$ and vanishes (up to logarithmic corrections) for the
infinite range Ising model.

We demonstrated that the critical slowing down of the flat-histogram
simulations can be overcome by either changing the representation of the
system to allow cluster updates or by changing the simulated ensemble and
keeping local updates. In order to apply cluster updates to continuous spin
models in connection with broad-histogram simulations, we introduced a new
spin-multigraph based representation. The broad-histogram simulations which
take advantage of cluster updates or an optimized statistical ensemble do not
suffer from critical slowing down and show optimal scaling in comparison to
an unbiased Markovian random walk.

\begin{acknowledgments}
We acknowledge financial support of the Swiss National Science Foundation
and the U.S.\ National Science Foundation DMR-0242402 (Machta), and NSF
DBI-0320875 (Colonna-Romano and Gould). We thank Sid Redner for useful
discussions.
\end{acknowledgments}


\begin{thebibliography}{99}
\bibitem{footnote1}
In Ref.~\cite{tt} the exact density of states was found using the Mathematica 
program of Beale~\cite{Beale}; here we used a C program and an 
infinite precision integer arithmetic package that also implements
Beale's approach. The latter approach allowed most of the 
calculation of the density of states to be spread over many processors.
\end{thebibliography}
\end{document}